\documentclass[twocolumn,prl,reprint,amsmath,amssymb,showpacs,
superscriptaddress,longbibliography]{revtex4-2}

\usepackage{CJK}
\usepackage{graphicx}
\usepackage{dcolumn}
\usepackage{bm}
\usepackage{graphicx}
\usepackage{epsfig}
\usepackage{epsf}
\usepackage{amssymb}
\usepackage{amsmath}
\usepackage{amsthm}
\usepackage{multirow}
\usepackage{verbatim}
\usepackage{cases}
\usepackage[colorlinks=true,linkcolor=blue,citecolor=blue,pdfauthor={ },pdftitle={ },pdfsubject={ },pdfkeywords={ }]{hyperref}
\usepackage{pgfplots}
\usepackage{float}
\usepackage{lipsum}
\usepackage{hyperref}
\usepackage{gensymb}

\begin{document}

\begin{CJK*}{UTF8}{gbsn}

\title{Identity Test of Single NV$^-$ Centers in Diamond at Hz-Precision Level}

\author{Tianyu Xie}
\thanks{These authors contributed equally to this work.}
\affiliation{Hefei National Laboratory for Physical Sciences at the Microscale, University of Science and Technology of China, Hefei 230026, China}
\affiliation{CAS Key Laboratory of Microscale Magnetic Resonance and Department of Modern Physics, University of Science and Technology of China, Hefei 230026, China}
\affiliation{Synergetic Innovation Center of Quantum Information and Quantum Physics, University of Science and Technology of China, Hefei 230026, China}

\author{Zhiyuan Zhao}
\thanks{These authors contributed equally to this work.}
\affiliation{Hefei National Laboratory for Physical Sciences at the Microscale, University of Science and Technology of China, Hefei 230026, China}
\affiliation{CAS Key Laboratory of Microscale Magnetic Resonance and Department of Modern Physics, University of Science and Technology of China, Hefei 230026, China}
\affiliation{Synergetic Innovation Center of Quantum Information and Quantum Physics, University of Science and Technology of China, Hefei 230026, China}

\author{Maosen Guo}
\thanks{These authors contributed equally to this work.}
\affiliation{Hefei National Laboratory for Physical Sciences at the Microscale, University of Science and Technology of China, Hefei 230026, China}
\affiliation{CAS Key Laboratory of Microscale Magnetic Resonance and Department of Modern Physics,  University of Science and Technology of China, Hefei 230026, China}
\affiliation{Synergetic Innovation Center of Quantum Information and Quantum Physics, University of Science and Technology of China, Hefei 230026, China}

\author{Mengqi Wang}
\affiliation{Hefei National Laboratory for Physical Sciences at the Microscale, University of Science and Technology of China, Hefei 230026, China}
\affiliation{CAS Key Laboratory of Microscale Magnetic Resonance and Department of Modern Physics, University of Science and Technology of China, Hefei 230026, China}
\affiliation{Synergetic Innovation Center of Quantum Information and Quantum Physics, University of Science and Technology of China, Hefei 230026, China}

\author{Fazhan Shi}
\email{fzshi@ustc.edu.cn}
\affiliation{Hefei National Laboratory for Physical Sciences at the Microscale, University of Science and Technology of China, Hefei 230026, China}
\affiliation{CAS Key Laboratory of Microscale Magnetic Resonance and Department of Modern Physics, University of Science and Technology of China, Hefei 230026, China}
\affiliation{Synergetic Innovation Center of Quantum Information and Quantum Physics, University of Science and Technology of China, Hefei 230026, China}

\author{Jiangfeng Du}
\email{djf@ustc.edu.cn}
\affiliation{Hefei National Laboratory for Physical Sciences at the Microscale, University of Science and Technology of China, Hefei 230026, China}
\affiliation{CAS Key Laboratory of Microscale Magnetic Resonance and Department of Modern Physics, University of Science and Technology of China, Hefei 230026, China}
\affiliation{Synergetic Innovation Center of Quantum Information and Quantum Physics, University of Science and Technology of China, Hefei 230026, China}

\begin{abstract}
Atomic-like defects in solids are not considered to be identical owing to the imperfections of host lattice. Here, we found that even under ambient conditions, negatively charged nitrogen-vacancy (NV$^-$) centers in diamond {could still} manifest identical at Hz-precision level, corresponding to a 10$^{-7}$-level relative precision, {while the lattice strain can destroy the identity by tens of Hz}. All parameters involved in the NV$^-$-$^{14}$N Hamiltonian are determined by formulating six nuclear frequencies at 10-mHz-level precision and measuring them at Hz-level precision. The most precisely measured parameter, the $^{14}$N quadrupole coupling $P$, is given by -4945754.9(8) Hz, whose precision is improved by nearly four orders of magnitude compared with previous measurements. We offer an approach for performing precision measurements in solids and deepening our understandings of NV centers as well as other solid-state defects. Besides, these high-precision results imply a potential application of a robust and integrated atomic-like clock based on ensemble NV centers.
\end{abstract}

\maketitle

\section{Introduction}
Precision measurements play a crucial role in many important fields including precision tests of the fundamental laws of physics \cite{new_physics_2018, LIGO_correlation_2020}, measurements of fundamental constants \cite{G_2018, alpha_2018}, atomic clocks \cite{lattice_clock_2014, lattice_clock_2019, ion_clock_2010, ion_clock_2019} and gravitational wave detection \cite{LIGO_2013, GW_2016}. The particles (electrons, atoms, and photons) used therein are naturally identical, which underlies these precision measurements. However, it seems a plausible common intuition that atomic-like defects in solids are not identical because of the complexity of host lattice. Besides, it is extremely difficult for measurements and theoretical analyses in a solid to be precise enough, especially under ambient conditions. Therefore, the identity test of atomic-like defects in solids with high precision remains elusive.

Over the last decade, a novel atomic-like defect in diamond, namely nitrogen-vacancy (NV) centers \cite{NV_2013}, has attracted widespread attention and been studied extensively. It is characteristic of many excellent properties: long-lived spin coherence even under ambient conditions \cite{coherence_2019}, high spin polarization by laser illumination \cite{SQL_2021}, optical readout of spin states and strongly coupled nuclear spins surrounding NV centers as quantum resources \cite{SQL_2021}. With these advantages, NV centers find a whole wealth of applications ranging from quantum sensing \cite{sensing_2017, sensitivity_2020} including magnetic resonance of single molecules \cite{ESR_2015, NMR_2016} and nanoscale magnetic imaging \cite{periodic_2017, hopping_2014}, to quantum computation \cite{error_correction_2014} and networks \cite{network_2017}.

In this Letter, we report on the first identity test in solids at Hz-precision level by employing single NV$^-$ centers in diamond under ambient conditions. Analytical formulas associating the parameters of the NV$^-$-$^{14}$N Hamiltonian with six nuclear frequencies are constructed with 10 mHz precision by analogy with coherent stimulated Raman transitions (CSRTs) \cite{QIP_ion_2003}. Six nuclear frequencies are measured with Hz precision by adopting Ramsey interferometry used in atomic clocks \cite{lattice_clock_2014, lattice_clock_2019, ion_clock_2010, ion_clock_2019}. Combined with two transition frequencies of the NV$^-$ spin, all parameters of the entire Hamiltonian can be solved out by least squares regression. We applied the method of measurement and analysis to multiple NV centers. It turns out that, most strikingly, five NV centers {(within a 60$\times$10 \textmu m$^2$ area)} far away from solid immersion lenses (SILs) appear identical at Hz-precision level while two NV centers inside a SIL differ by tens of Hz. By synthesizing the experimental data of seven NV centers, four key parameters, i.e., the $^{14}$N quadrupole coupling $P$, the longitudinal component $A_{\parallel}$, the transverse component $A_{\perp}$ of the hyperfine interaction, and the ratio $\gamma_e/\gamma_n$ of the gyromagnetic ratio of the NV$^-$ electron to that of $^{14}$N, are given by -4945754.9(8) Hz, -2164689.8(1.2) Hz, -2632.7(4) kHz and -9113.85(4), respectively. The values in parentheses stand for one standard deviation. Compared with previous experiments \cite{NV_2013, nuclear_2009, decoherence_protected_2012, transverse_2015}, the uncertainties of the first two values are reduced by more than three orders of magnitude, while the last two nearly two orders of magnitude (see Supplemental Material \cite{SM}).

\section{System and theoretical model}
\begin{figure}
\includegraphics[width=1\columnwidth]{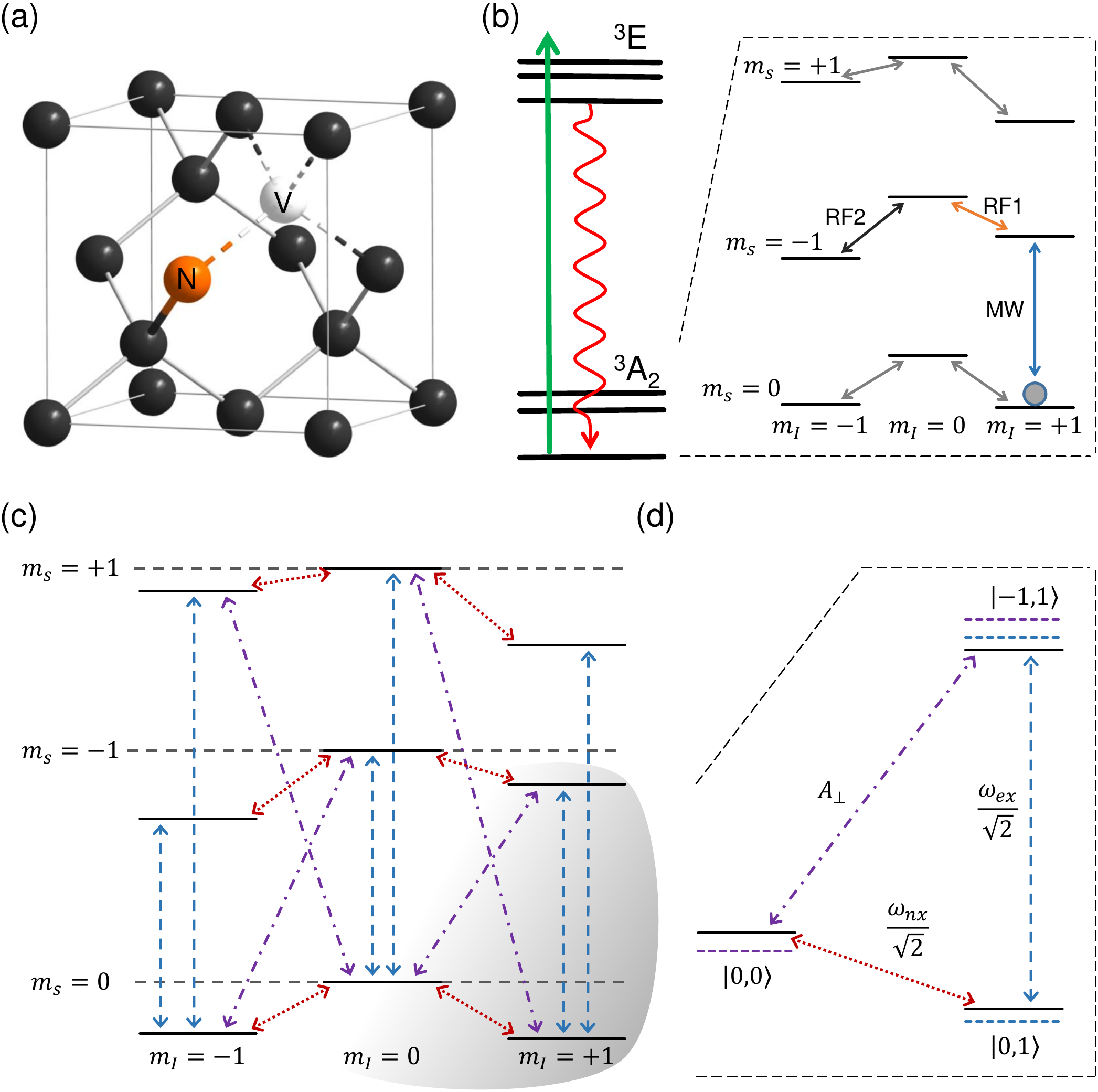}
\caption{Characterization of the NV center coupled with the attendant $^{14}$N nuclear spin and its spin states in the ground state $^{3}A_2$. (a) Atomic structure of the NV center in diamond lattice. The orange sphere denotes the nitrogen atom while the white one the vacancy and the black one carbon atoms. (b) Level diagrams of the NV$^-$ electron and the coupled electron and nuclear spins in the ground state. The state marked by the gray circle is the initialized state. The Gray, black, and orange arrows indicate the six nuclear spin transitions driven by RF pulses. The blue arrow shows the electron spin transition driven by MW pulses. The MW, RF1, and RF2 pulses are used in the pulse sequence in Fig. \ref{experiments}(a). (c) Level structure of the NV$^-$-$^{14}$N system described by the principal term $H_{\parallel}$ of the entire Hamiltonian. The lines with two arrowheads denote the transition matrix elements from the perturbative term $H_{\perp}$. The dashdotted (purple), dashed (blue), and dotted (red) lines stand for the transverse hyperfine interaction, the transverse Zeeman terms for the electron spin and the nuclear spin in $H_{\perp}$, respectively. (d) Reduced three-level system in analogy to CSRT. Energy level repulsions induced by the transverse hyperfine interaction and the transverse Zeeman term of the electron spin are displayed by dashdotted and dashed lines. The values are the transition matrix elements of three transverse items.
}
\label{system}
\end{figure}

The NV$^-$ center in diamond with electron spin $S=1$ in the ground state consists of a substitutional $^{14}$N atom with nuclear spin $I=1$ and an adjacent vacancy as shown in Fig. \ref{system}(a). Here we focus on the ground state ${^3}A_2$ of the NV$^-$ electron with long-lived spin coherence. Under a magnetic field $B_0 \approx 510$ G aligned along the NV axis, both the electron spin and the nuclear spin are jointly initialized by a 532-nm laser pulse \cite{pol_2009} in Fig. \ref{system}(b). Microwave (MW) and radio-frequency (RF) pulses are used to coherently manipulate the electron spin and the nuclear spin. The spin states can be read out by collecting fluorescence photons after laser illumination based on spin-dependent intersystem crossing (ISC) \cite{NV_2013}.

Two kinds of interactions, i.e., the nuclear quadrupole coupling of the $^{14}$N nucleus and the hyperfine interaction between the NV$^-$ electron spin and the nuclear spin, are used here to test the identity of the electron wavefunction $\Psi_e(\mathbf{r})$ of the NV$^-$ center. The wavefunction has $C_{3v}$ symmetry along the NV axis, defined as the z axis. The symmetry simplifies the $^{14}$N nuclear quadrupole coupling $P$ induced by the electric field gradient from the electron wavefunction interacting with the electric quadruple moment $Q$ of the $^{14}$N nucleus, which reads \cite{quad_1973}
\begin{equation}
\label{quadrupole_interaction}
H_{\rm{quad}} = \frac{3Q}{4} I_z^2 \bigg\langle \frac{\partial^2 V(\mathbf{r})}{\partial z^2} \bigg\rangle_{\mathbf{r}=\mathbf{r}_{\rm{N}}}=P I_z^2 
\end{equation}
where $I_z$ is the z-component of the nuclear spin operators, $\langle\cdot\rangle$ denotes an average over the wavefunction $\Psi_e(\mathbf{r})$, and $V(\mathbf{r})$ is the electrostatic potential produced by the NV$^-$ electron at point $\mathbf{r}$. $\langle \frac{\partial^2 V(\mathbf{r})}{\partial z^2} \rangle_{\mathbf{r}=\mathbf{r}_{\rm{N}}}$ denotes the zz-component of the electric field gradient at the position $\mathbf{r}_{\rm{N}}$ of the $^{14}$N nucleus. Likewise, the hyperfine interaction is constrained by the $C_{3v}$ symmetry. Two parameters, the longitudinal component $A_{\parallel}$ and the transverse component $A_{\perp}$, are adequate to describe the hyperfine interaction \cite{hyperfine_1998}
\begin{equation}
\label{hyperfine_interaction}
\begin{split}
H_{\rm{hyper}} &= -\frac{\mu_0 \gamma_e\gamma_n \hbar^2}{4\pi} \bigg[ \bigg\langle \frac{3(\mathbf{S} \cdot \mathbf{\hat{n}})(\mathbf{I} \cdot \mathbf{\hat{n}})-\mathbf{S} \cdot \mathbf{I}}{|\mathbf{r}-\mathbf{r}_{\rm{N}}|^3} \bigg\rangle\\
&\quad\,+\frac{8\pi |\Psi_e(\mathbf{r}_{\rm{N}})|^2}{3}\mathbf{S} \cdot \mathbf{I} \bigg]\\
&= A_{\parallel} S_z I_z+A_{\perp}(S_x I_x+S_y I_y)
\end{split}
\end{equation}
\noindent{where $\mu_0$ is the vacuum permeability and $\hbar$ is the reduced Planck constant. $\gamma_e$ and $\gamma_n$ are the gyromagnetic ratios of the electron spin and the $^{14}$N nuclear spin, and $\mathbf{S}=(S_x,S_y,S_z)$ and $\mathbf{I}=(I_x,I_y,I_z)$ are the spin operator vectors of two spins, respectively. $\mathbf{\hat{n}}$ is the unit vector along the direction of $\mathbf{r}-\mathbf{r}_{\rm{N}}$.} $|\Psi_e(\mathbf{r}_{\rm{N}})|^2$ the electron spin density at the $^{14}$N site. From Eq. (\ref{quadrupole_interaction}) and (\ref{hyperfine_interaction}), the parameters $P$, $A_{\parallel}$, and $A_{\perp}$ are all weighted averages over the electron wavefunction $\Psi_e(\mathbf{r})$, and thus good measures of the identity of NV$^-$ centers.

With the zero-field splitting of the electron spin, Zeeman effects of the electron spin and the nuclear spin included, the entire Hamiltonian is given by
\begin{gather}
H_0 = H_{\parallel}+H_{\perp},\label{Hamiltonian}\\
H_{\parallel} = D S_z^2+\omega_{e}S_z+P I_z^2+\omega_{n}I_z+A_{\parallel} S_z I_z,\label{principal}\\
H_{\perp} = A_{\perp}(S_x I_x+S_y I_y)+\omega_{ex}S_x+\omega_{nx}I_x,\label{perturbative}
\end{gather}
where $H_{\parallel}$ is the principal term of the Hamiltonian, and $H_{\perp}$ is the perturbative term that is non-commutable with the principal term. $D\approx 2870$ MHz is the zero-field splitting of the electron spin. $\omega_{e}$ and $\omega_{ex}$ are the longitudinal component and the transverse component of the Zeeman effect of the electron spin, while $\omega_{n}$ and $\omega_{nx}$ the nuclear spin. The alignment of the magnetic field with the NV axis is adjusted via the method of monitoring the counts of fluorescence photons \cite{pol_2009}. The slight misalignment should be taken into consideration as $\omega_{ex}$ and $\omega_{nx}$ in view of desired high-precision measurements. {The strain effect on the electron spin \cite{strain_2018} is ignored here due to the vast energy gaps of the electron spin, and a detailed analysis is included in the Supplemental Material \cite{SM}.} Fig. \ref{system}(c) depicts the level structure of the coupled electron and nuclear spins obtained from $H_{\parallel}$. The perturbative term $H_{\perp}$ generates transition matrix elements to mix two states and induces small shifts of energy levels. Approximating the energy levels by second-order perturbation could reduce this nine-level system into multiple three-level systems detailed in Fig. \ref{system}(d). The three-level hamiltonian appears in a general form
\begin{equation}
\begin{pmatrix}
\Delta & a & b\\
a & \delta_1 & c\\
b & c & \delta_2 
\end{pmatrix}
\end{equation}
where $a$, $b$, and $c$ are transition matrix elements and $\Delta$, $\delta_1$, and $\delta_2$ are energy levels with $\Delta \gg \delta_1, \delta_2$. Performing the approximation like that of CSRT \cite{QIP_ion_2003} with a small but significant modification gives
\begin{equation}
\begin{pmatrix}
\Delta + \frac{a^2}{\Delta-\delta_1} + \frac{b^2}{\Delta-\delta_2} & 0 & 0\\
0 & \delta_1 - \frac{a^2}{\Delta-\delta_1} & c-\frac{a b}{\Delta}\\
0 & c-\frac{a b}{\Delta} & \delta_2 - \frac{b^2}{\Delta-\delta_2} 
\end{pmatrix}
\end{equation}
The small modification is inserting $\delta_1$ and $\delta_2$ into the denominators of diagonal elements accordingly. Without this modification, the final formulas of six nuclear frequencies only have 10 Hz precision. Multiple three-level systems take effect together via multipath interference, and then no transition matrix elements exist between the electron spin states. By performing again the second-order perturbation for the nuclear spin in each subspace of the electron spin, we deduce six analytical formulas for six nuclear transition frequencies. Compared with the results of numerical simulation, it comes out that the formulas have 10 mHz precision (see Supplemental Material \cite{SM}).

\section{Experimental results}
\begin{figure}
\includegraphics[width=1\columnwidth]{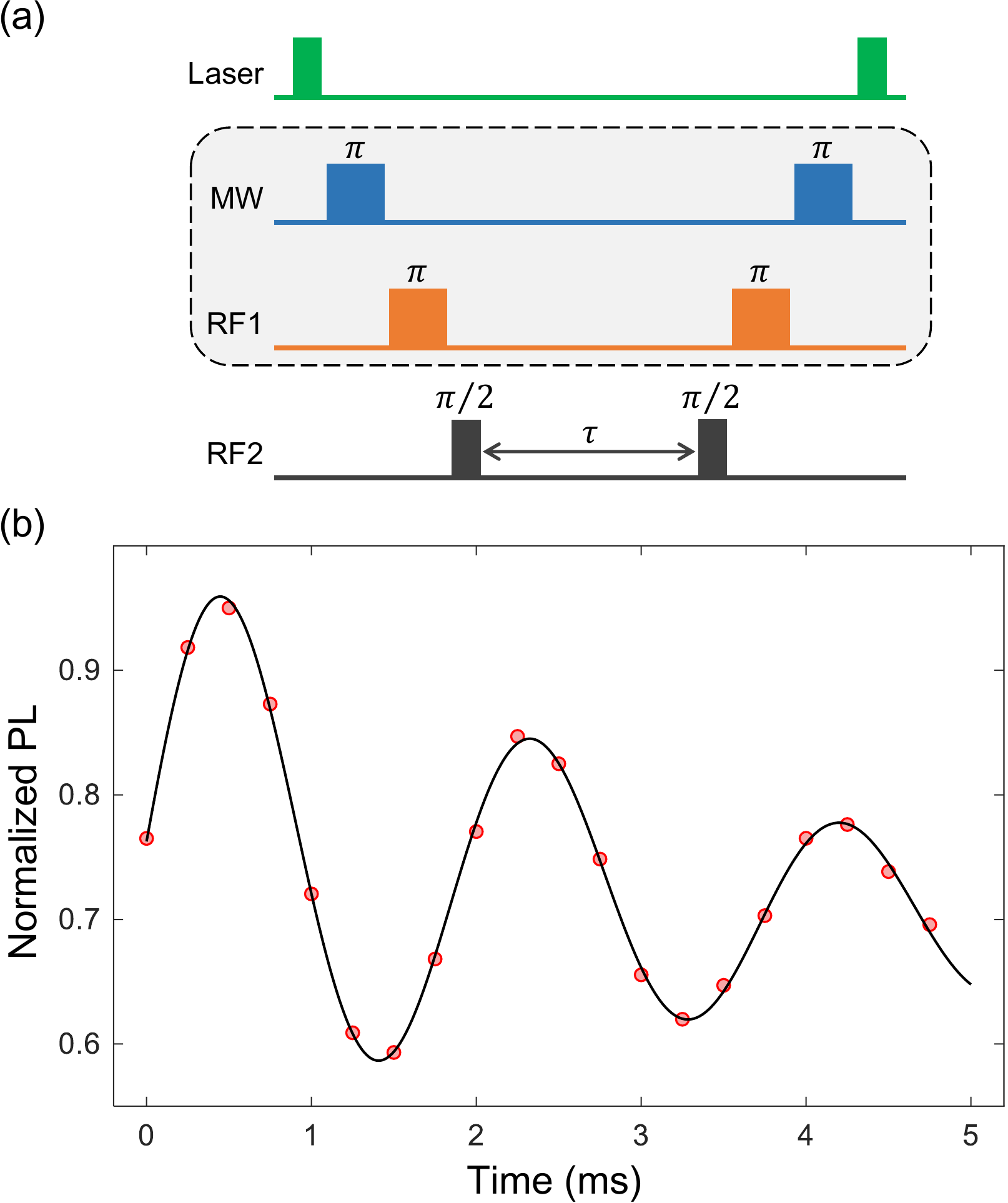}
\caption{Ramsey interferometry and interference pattern. (a) The pulse sequence of laser, MW, and RF for Ramsey interference between the state $|m_S=-1,m_I=0\rangle$ and the state $|m_S=-1,m_I=-1\rangle$. The sequence encircled by the dashed box is aimed at population transfer to the state for performing Ramsey interference. Find the corresponding manipulation of spin states in Fig. \ref{system}(b) according to the names or the colors. (b) Resulting interference pattern after applying the sequence above to the last NV center in Fig. \ref{identity}. The black line is data fitting using the function $[a\sin(2\pi (\delta f) t+\phi_0)+b]e^{-(t/T_2^*)^p}+c$ with $\delta f$ meaning the detuning.
}
\label{experiments}
\end{figure}

\begin{figure}
\includegraphics[width=1\columnwidth]{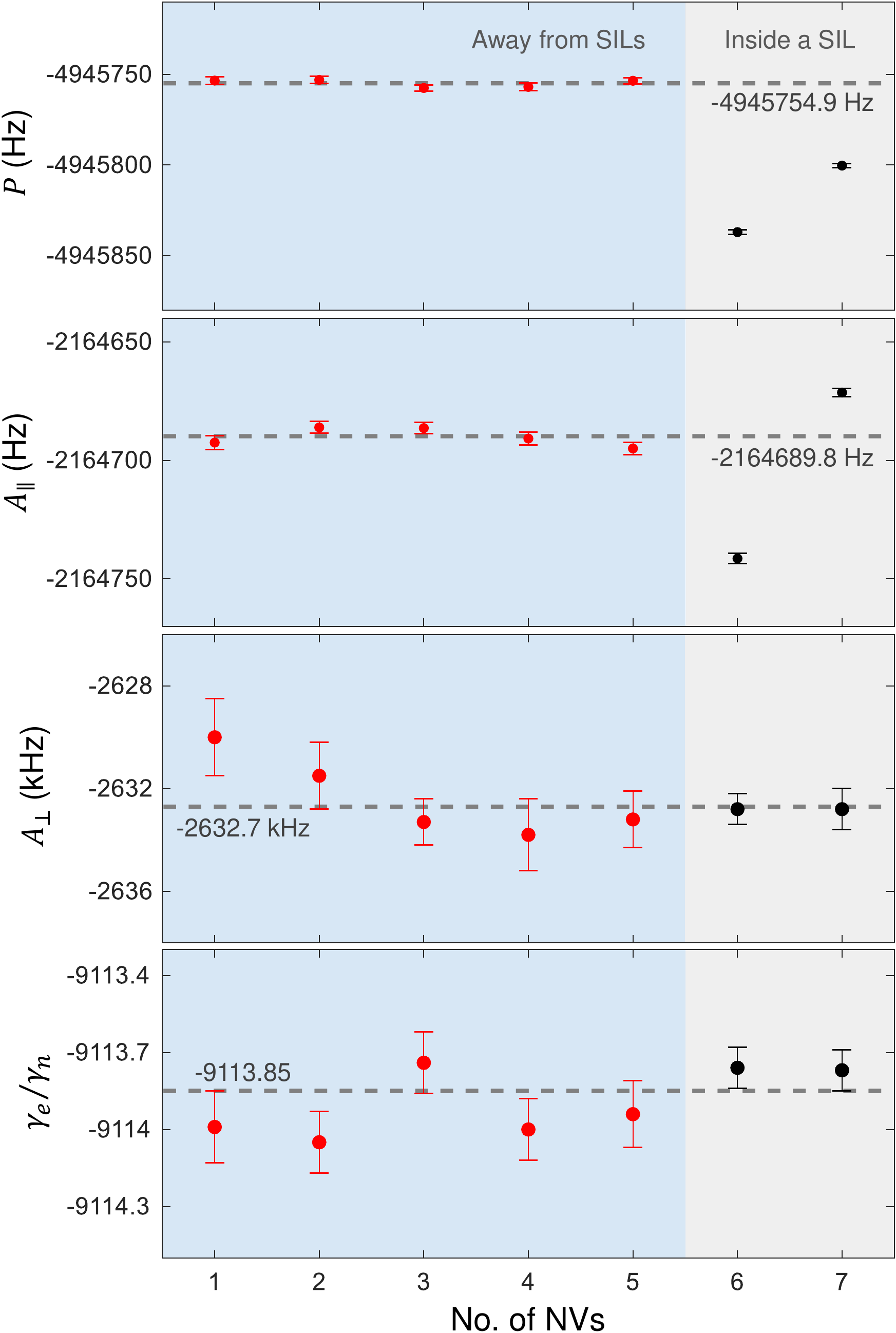}
\caption{Identity test. From top to bottom: the $^{14}$N quadrupole coupling $P$, the longitudinal component $A_{\parallel}$, the transverse component $A_{\perp}$ of the hyperfine interaction, and the ratio $\gamma_e/\gamma_n$ of the gyromagnetic ratio of the NV$^-$ electron to that of $^{14}$N for seven NV centers. Among them, the first five NVs (red) are far away from SILs and the last two NVs (black) are inside a SIL. The values represented by the dashed lines are the weighted averages of the NVs far away from SILs for $P$ and $A_{\parallel}$, and all seven NVs for $A_{\perp}$ and $\gamma_e/\gamma_n$.
}
\label{identity}
\end{figure}

There are seven independent parameters in total in Eq. (\ref{principal}) and (\ref{perturbative}) since $\omega_{ex}/\omega_e$ equals $\omega_{nx}/\omega_n$. Two MW frequencies for the electron spin determine the values of $D$ and $\omega_e$. The left five parameters can be determined by six nuclear frequencies in three subspaces of the electron spin via least squares regression. The statistical errors can be derived by the method of error propagation where the derivatives associating five parameters and six nuclear frequencies are numerically calculated out (see Supplemental Material \cite{SM}).

In order to measure transition frequencies with high precision, we adopted the measurement scheme of Ramsey interferometry, commonly used in atomic clocks \cite{lattice_clock_2014, lattice_clock_2019, ion_clock_2010, ion_clock_2019}. Fig. \ref{experiments}(a) is an examplary pulse sequence for measuring the transition frequency between the state $|m_S=-1,m_I=0\rangle$ and the state $|m_S=-1,m_I=-1\rangle$. The sequence for population transfer should be adapted accordingly for the measurements of other nuclear transtions. The resulting interference pattern is depicted by Fig. \ref{experiments}(b) with decoherence. The overall decline of the pattern is caused by the longitudinal relaxation of the electron spin. By fitting the data, the detuning $\delta f$ is estimated to be 533.2(1.6) Hz. By adding the detuning $\delta f$ to the RF2 frequency in Fig. \ref{experiments}(a), we acquired the value of -6958568.8(1.6) Hz for the transition between the state $|m_S=-1,m_I=0\rangle$ and the state $|m_S=-1,m_I=-1\rangle$. The fluctuation of the magnetic field is stabilized below 0.3 \textmu T to ensure the Hz-precision measurement (see Supplemental Material \cite{SM}). The same method is applied to the other five transitions, and then the left five parameters are determined by the data processing described above with a Hz-level residual. We found that abandoning the parameter $\omega_{ex}$ during the data processing still produces a Hz-level residual, which signifies that the magnetic field is aligned parallel enough ($<0.1\degree$) to the NV axis (see Supplemental Material \cite{SM}). Therefore, the transverse Zeeman term of the electron spin $\omega_{ex}$ will be ignored in what follows. Besides, the ratio of two Zeeman shifts $\omega_e/\omega_n=\gamma_e/\gamma_n$ are independent of the magnetic field and an intrinsic property of NV centers as another indicator of the identity test together with $P$, $A_{\parallel}$, and $A_{\perp}$.

To test the identity of NV$^-$ centers, we measured the parameters of seven NV centers, as displayed in Fig. \ref{identity}. Among them, five NV centers {are randomly chosen far away from all SILs and distributed within a 60$\times$10 \textmu m$^2$ area,} and two NV centers are inside a SIL. In terms of $P$ and $A_{\parallel}$, most strikingly, five NVs away from SILs are identical at Hz-precision level within two standard deviations. It corresponds to a $10^{-7}$-level relative precision for the quadrupole coupling $P$. But two NVs inside the same SIL differ slightly by tens of Hz beyond tens of standard deviations. The fact that the strain generated during creating SILs breaks the symmetry of the diamond lattice is responsible for the observed differences. {Morever, through the identity test of ten more NV centers away from SILs, we found that the intrinsic strain in diamond can also take effect and destroy the identity of NV$^-$ centers by tens of Hz (see Supplemental Material \cite{SM}).} A quantitative description of theory for the phenomenon is urgently needed. The results imply a possibility for studying the susceptibility of NV$^-$ centers to strain \cite{strain_2014, strain_2018} as well as two other relevant quantities, temperature \cite{temp_2010, temp_2019} and electric field \cite{electric_1990, electric_2019}. As for $A_{\perp}$ and $\gamma_e/\gamma_n$, no evident differences are observed for all seven centers due to relatively low precisions compared with $P$ and $A_{\parallel}$. By weightedly averaging all available parameter values, $P$, $A_{\parallel}$, $A_{\perp}$, and $\gamma_e/\gamma_n$ are given by -4945754.9(8) Hz, -2164689.8(1.2) Hz, -2632.7(4) kHz and -9113.85(4).

\section{Discussion}
\begin{figure}
\includegraphics[width=1\columnwidth]{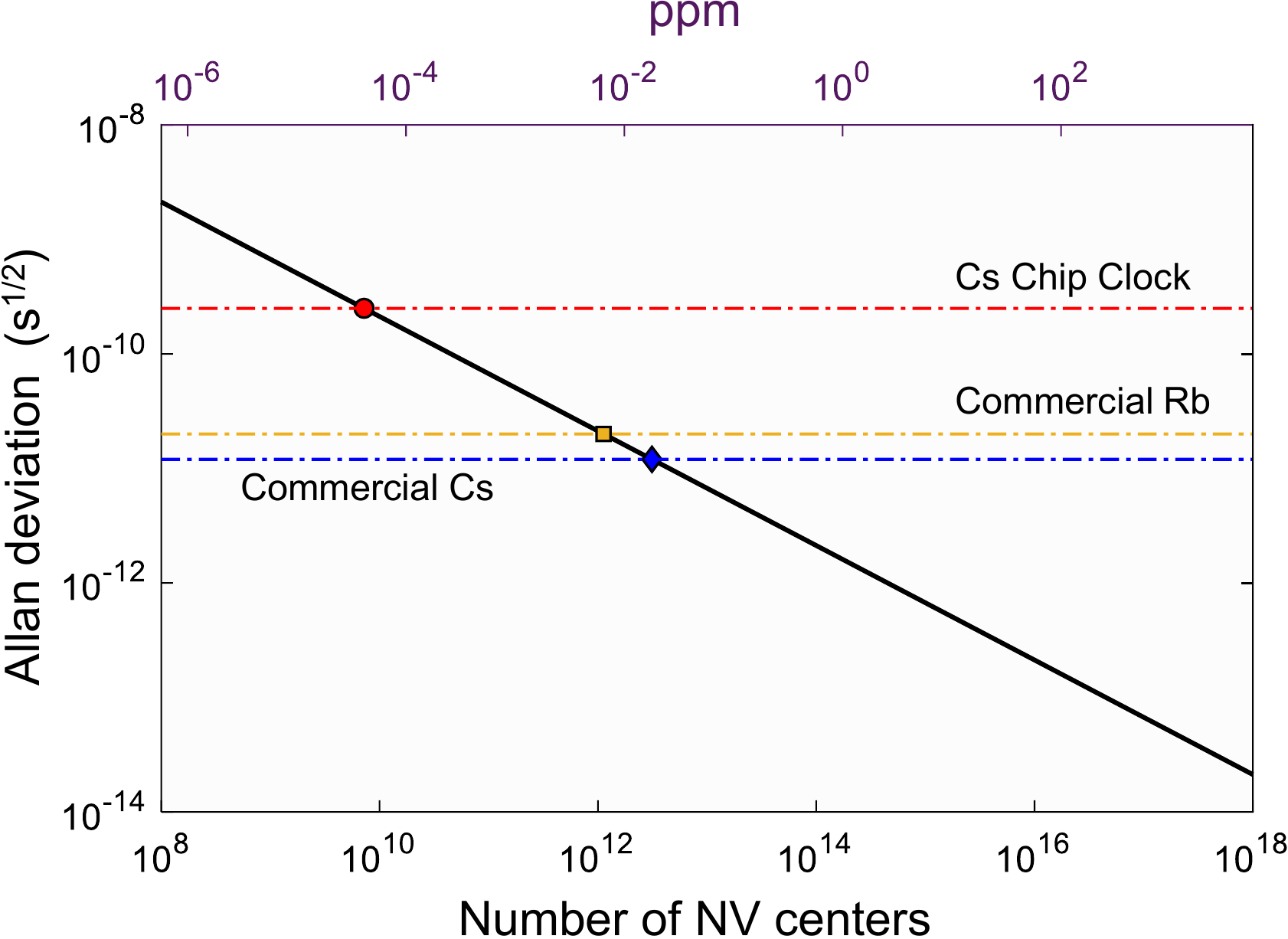}
\caption{Allan deviation of the proposed NV clock. The solid black line shows the fractional frequency instability at 1 s of integration as a function of the number of ensemble NV centers (bottom) and the density of NV centers (top) for a 1 mm$^3$ diamond. For comparison, the fractional frequency instability is $2.5\times 10^{-10}$ for Cs chip clock \cite{chip_clock_2004}, $2\times 10^{-11}$ for commercial Rb clock \cite{Rb_clock}, and $1.2\times 10^{-11}$ for commercial Cs clock \cite{Cs_clock}.
}
\label{clock}
\end{figure}

One kind of atomic-like clock based on NV$^-$ centers \cite{D_clock_2013} has been proposed by measuring the zero-field splitting $D$ of the electron spin. However, the frequency stability is subject to the temperature fluctuation with a coefficient of -74 kHz/K \cite{temp_2010}. As a result, it is not suitable for commercial applications. Based on our high-precision results, here we propose a new scheme by measuring the transition frequencies of the $^{14}$N nuclear spin dominated by the quadrupole coupling $P$. The temperature susceptibility of the hyperfine interaction has been studied recently in \cite{temp_2019}. No evident shift is observed across hundreds of Kelvin but it still awaits a higher-precision measurement. The magnetic susceptibility can be overcome by directly removing the external magnetic field and shielding the magnetic field from the environment. An ensemble of NV centers is utilized to reduce the statistical error of the frequency measurement \cite{sensitivity_2020}. The fractional frequency instability for an ensemble of $N$ NV centers with a measurement time $T$ is calculated by
\begin{equation}
\frac{\delta f}{f_0}=\frac{1}{2\pi f_0 F\sqrt{T_2^* T}\sqrt{N}}\sim \frac{2\times 10^{-5}}{\sqrt{T}\sqrt{N}}
\label{instability_clock}
\end{equation}
where $f_0$ is dominated by $P\sim5$ MHz, $F\sim1.5\%$ is the readout fidelity of NV centers, and $T_2^*\sim10$ ms is the coherence time of the nuclear spin. The fractional frequency instability at 1 s of integration $\delta f\sqrt{T}/f_0$ is plotted in Fig. \ref{clock} in comparison with three commercial atomic clocks. An ensemble of $10^{12}$ NV centers is enough to achieve the level of commercial clocks, corresponding to a density of $\sim6$ ppb for a 1 mm$^3$ diamond. Compared with commercial clocks, the clock based on NV centers are more robust and integratable because NV centers are trapped in a strong diamond lattice. They are extremely suitable for some challenging environments such as cryogenic environment, high pressure, moving platforms, and so forth \cite{challenging_2020}. By using a bigger diamond with properly denser NV centers ($\sim100$ ppm is achievable in \cite{decoherence_2020}), the clock proposed here may outperform commercial atomic clocks by several orders of magnitude.

{The impurities surrounding the NV center may shift $P$ and $A_{\parallel}$ through the electric field effect, of which the most abundant one is nitrogen. The nitrogen concentration of the diamond we used here is less than 5 ppb and thus has no effect on $P$ and $A_{\parallel}$. The ensemble NV centers with a nitrogen density of $\sim1$ ppm may have a distribution of several Hz for $P$ and $A_{\parallel}$ (see Supplemental Material \cite{SM}). The deviation at this level has no effect on the coherence time of the nuclear spin $T_2^*\sim10$ ms and thus will not lower the performance of the proposed atomic-like clock in Eq. (\ref{instability_clock}).}

\section{Conclusion}
In summary, we measured six transtion frequencies of $^{14}$N nuclear spin with Hz precision under ambient conditions by means of Ramsey interferometry. Through an analysis and calculation with 10 mHz precision, the values of four key parameters in the Hamiltonian, $P$, $A_{\parallel}$, $A_{\perp}$, and $\gamma_e/\gamma_n$, are obtained to be -4945754.9(8) Hz, -2164689.8(1.2) Hz, -2632.7(4) kHz and -9113.85(4). These results have direct benefits for high-fidelity quantum control and computation \cite{error_correction_2014, network_2017, control_2015}. By comparing the parameters $P$ and $A_{\parallel}$ of seven NV centers, we found that five NV centers {(within a 60$\times$10 \textmu m$^2$ area)} far away from SILs are identical at Hz-precision level and two NV centers inside a SIL differ by tens of Hz. It is a high-precision approach for studying the susceptibilities of many physical quantities including stain \cite{strain_2014, strain_2018}, temperature \cite{temp_2010, temp_2019}, and electric field \cite{electric_1990, electric_2019}. Combined with theoretical researches, it perhaps enhances our understanding of the basic physics of NV centers. In the future, this kind of test could be improved to mHz-precision level, which is performed on an isotopically purified diamond at cryogenic temperature. Finally, it is expected to construct a robust and integrated atomic-like clock using ensemble NV centers with a better performance than commercial atomic clocks nowadays.

\section*{Acknowledgments}
This work was supported by the National Natural Science Foundation of China (Grant Nos. 91636217, 81788101, 11722544, 11761131011), the National Key R$\&$D Program of China (Grant Nos. 2018YFA0306600 and 2016YFA0502400), the CAS (Grant Nos. GJJSTD20170001 and QYZDY-SSW-SLH004), the Anhui Initiative in Quantum Information Technologies (Grant No. AHY050000), and the Fundamental Research Funds for the Central Universities.

\end{CJK*}

\end{document}


\begin{CJK*}{UTF8}{gbsn}

\title{Supplemental Material for \\ Identity Test of Single NV$^-$ Centers in Diamond at Hz-Precision Level}

\author{Tianyu Xie}
\thanks{These authors contributed equally to this work.}
\affiliation{Hefei National Laboratory for Physical Sciences at the Microscale, University of Science and Technology of China, Hefei 230026, China}
\affiliation{CAS Key Laboratory of Microscale Magnetic Resonance and Department of Modern Physics, University of Science and Technology of China, Hefei 230026, China}
\affiliation{Synergetic Innovation Center of Quantum Information and Quantum Physics, University of Science and Technology of China, Hefei 230026, China}

\author{Zhiyuan Zhao}
\thanks{These authors contributed equally to this work.}
\affiliation{Hefei National Laboratory for Physical Sciences at the Microscale, University of Science and Technology of China, Hefei 230026, China}
\affiliation{CAS Key Laboratory of Microscale Magnetic Resonance and Department of Modern Physics, University of Science and Technology of China, Hefei 230026, China}
\affiliation{Synergetic Innovation Center of Quantum Information and Quantum Physics, University of Science and Technology of China, Hefei 230026, China}

\author{Maosen Guo}
\thanks{These authors contributed equally to this work.}
\affiliation{Hefei National Laboratory for Physical Sciences at the Microscale, University of Science and Technology of China, Hefei 230026, China}
\affiliation{CAS Key Laboratory of Microscale Magnetic Resonance and Department of Modern Physics,  University of Science and Technology of China, Hefei 230026, China}
\affiliation{Synergetic Innovation Center of Quantum Information and Quantum Physics, University of Science and Technology of China, Hefei 230026, China}

\author{Mengqi Wang}
\affiliation{Hefei National Laboratory for Physical Sciences at the Microscale, University of Science and Technology of China, Hefei 230026, China}
\affiliation{CAS Key Laboratory of Microscale Magnetic Resonance and Department of Modern Physics, University of Science and Technology of China, Hefei 230026, China}
\affiliation{Synergetic Innovation Center of Quantum Information and Quantum Physics, University of Science and Technology of China, Hefei 230026, China}

\author{Fazhan Shi}
\email{fzshi@ustc.edu.cn}
\affiliation{Hefei National Laboratory for Physical Sciences at the Microscale, University of Science and Technology of China, Hefei 230026, China}
\affiliation{CAS Key Laboratory of Microscale Magnetic Resonance and Department of Modern Physics, University of Science and Technology of China, Hefei 230026, China}
\affiliation{Synergetic Innovation Center of Quantum Information and Quantum Physics, University of Science and Technology of China, Hefei 230026, China}

\author{Jiangfeng Du}
\email{djf@ustc.edu.cn}
\affiliation{Hefei National Laboratory for Physical Sciences at the Microscale, University of Science and Technology of China, Hefei 230026, China}
\affiliation{CAS Key Laboratory of Microscale Magnetic Resonance and Department of Modern Physics, University of Science and Technology of China, Hefei 230026, China}
\affiliation{Synergetic Innovation Center of Quantum Information and Quantum Physics, University of Science and Technology of China, Hefei 230026, China}

\maketitle

\section{Detailed theoretical model}
\subsection{Detailed theoretical derivation}
As described in the text, the formulas of six transition frequencies of the $^{14}$N nuclear spin can be derived by multipath interference of multiple reduced three-level systems. The Hamiltonian of the three-level system shown in Fig. 1(d) is given by
\begin{equation}
\begin{pmatrix}
D-\omega_e+Q+\omega_n-A_{\parallel} & A_{\perp} & \omega_{ex}/\sqrt{2}\\
A_{\perp} & 0 & \omega_{nx}/\sqrt{2}\\
\omega_{ex}/\sqrt{2} & \omega_{nx}/\sqrt{2} & Q+\omega_n
\end{pmatrix}
\end{equation}
where the first state in the matrix above is $|m_S=-1,m_I=+1\rangle$, the second $|m_S=0,m_I=0\rangle$, and the third $|m_S=0,m_I=+1\rangle$. Performing the approximation from Eq. (6) to Eq. (7) in the text gives
\begin{equation}
\begin{psmallmatrix}
D-\omega_e+Q+\omega_n-A_{\parallel}+\frac{A_{\perp}^2}{D-\omega_e+Q+\omega_n-A_{\parallel}}+\frac{\omega_{ex}^2/2}{D-\omega_e-A_{\parallel}}
& 0 & 0\\
0 & -\frac{A_{\perp}^2}{D-\omega_e+Q+\omega_n-A_{\parallel}} & \frac{1}{\sqrt{2}}(\omega_{nx}-\frac{\omega_{ex}A_{\perp}}{D-\omega_e})\\
0 & \frac{1}{\sqrt{2}}(\omega_{nx}-\frac{\omega_{ex}A_{\perp}}{D-\omega_e}) & Q+\omega_n-\frac{\omega_{ex}^2/2}{D-\omega_e-A_{\parallel}}
\end{psmallmatrix}
\end{equation}
Implementing the approximation to all three-level systems makes the transition matrix elements between different subspaces of the NV electron spin zeros. Then through multipath interference, we acquire the Hamiltonian in the $m_s=0$ subspace of the electron spin
\begin{equation}
\begin{pmatrix}
\omega_{+1} & \Omega/\sqrt{2} & 0\\
\Omega/\sqrt{2} & \omega_0 & \Omega/\sqrt{2}\\
0 & \Omega/\sqrt{2} & \omega_{-1}
\end{pmatrix}
\label{subspace}
\end{equation}
with
\begin{gather}
\omega_{+1}=Q+\omega_n-\frac{A_{\perp}^2}{D+\omega_e-Q-\omega_n}-\frac{\omega_{ex}^2/2}{D-\omega_e-A_{\parallel}}-\frac{\omega_{ex}^2/2}{D+\omega_e+A_{\parallel}} \label{omega_+1}\\
\omega_{0}=-\frac{A_{\perp}^2}{D-\omega_e+Q+\omega_n-A_{\parallel}}-\frac{A_{\perp}^2}{D+\omega_e+Q-\omega_n-A_{\parallel}}-\frac{\omega_{ex}^2/2}{D-\omega_e}-\frac{\omega_{ex}^2/2}{D+\omega_e} \label{omega_0}\\
\omega_{-1}=Q-\omega_n-\frac{A_{\perp}^2}{D-\omega_e-Q+\omega_n}-\frac{\omega_{ex}^2/2}{D-\omega_e+A_{\parallel}}-\frac{\omega_{ex}^2/2}{D+\omega_e-A_{\parallel}} \label{omega_-1}\\
\Omega=\omega_{nx}-\frac{\omega_{ex}A_{\perp}}{D-\omega_e}-\frac{\omega_{ex}A_{\perp}}{D+\omega_e}
\end{gather}
Eq. (\ref{subspace}) can be approximated again by second-order perturbation
\begin{equation}
\begin{pmatrix}
\omega_{+1}-\frac{\Omega^2/2}{\omega_{0}-\omega_{+1}} & 0 & 0\\
0 & \omega_0+\frac{\Omega^2/2}{\omega_{0}-\omega_{+1}}+\frac{\Omega^2/2}{\omega_{0}-\omega_{-1}} & 0\\
0 & 0 & \omega_{-1}-\frac{\Omega^2/2}{\omega_{0}-\omega_{-1}}
\end{pmatrix}
\end{equation}
Therefore, two transition frequencies of the nuclear spin in the $m_S=0$ subspace are formulated as
\begin{gather}
f_{0,0\leftrightarrow +1}=\omega_0-\omega_{+1}+\frac{\Omega^2}{\omega_{0}-\omega_{+1}}+\frac{\Omega^2/2}{\omega_{0}-\omega_{-1}} \label{f_0to+1}\\
f_{0,0\leftrightarrow -1}=\omega_0-\omega_{-1}+\frac{\Omega^2}{\omega_{0}-\omega_{-1}}+\frac{\Omega^2/2}{\omega_{0}-\omega_{+1}} \label{f_0to-1}
\end{gather}
Similar formulas can be obtained for the transition frequencies in the $m_S=+1$ and $m_S=-1$ subspaces. Compared with numerical simulation of the original Hamiltonian, the formulas above have 10-mHz-level precision. If small terms, i.e, $Q$, $A_{\parallel}$, and $\omega_n$, are deleted from the denominators of Eq. (\ref{omega_+1}), (\ref{omega_0}), and (\ref{omega_-1}), the precisions of the formulas are decreased down to 10-Hz level.

\subsection{The strain effect on the electron spin}
The strain effect on the electron spin is ignored in the main text, and reconsidered here. The Hamiltonian allowed by the $C_{3v}$ symmetry of the NV center has the following form \cite{strain_2018}
\begin{equation}
H_{{\rm str}} = E_z S_z^2 + [E_x^{\prime} \{S_x,S_z\}+E_y^{\prime} \{S_y,S_z\}] + [E_x (S_y^2-S_x^2)+E_y \{S_x,S_y\}]
\label{strain}
\end{equation}
where $S_x$, $S_y$ and $S_z$ are the spin operators of the NV$^-$ electron spin. $E_z$, $E_x^{\prime}$, $E_y^{\prime}$, $E_x$ and $E_y$ are the strain parameters containing the strain tensor and spin-strain coupling parameters \cite{strain_2018}.  The first term just slightly changes the zero-field-splitting $D$ of the electron spin, and thus has been considered by measuring two electron transitions. The second term is non-commutable with the principal term $H_{\parallel}$ in the text and induces single quantum transitions between $m_S$=0 and $\pm 1$, which has a similar behavior with the transverse component of the Zeeman effect. The third term is also non-commutable with $H_{\parallel}$ and induces the double quantum transition between $m_S$=+1 and -1. These two items can be included in the derivation of the subsection above as well, like the transverse component of the Zeeman effect. But considering its smaller effect on six nuclear transitions, we just analyze it by numerical simulation. It turns out that the shifts of six nuclear transition frequencies will be less than 1 Hz if $E_x^{\prime}$, $E_y^{\prime}$, $E_x$ and $E_y$ are smaller than 1 MHz (for the NV centers away from SILs, these parameters are at the level of  $\sim$ kHz; for the NV centers in SILs, they are smaller than or at the level of $\sim$ 1 MHz). Therefore, the strain effect on six nuclear transitions is negligible. Besides, the least squares regressions with four parameters are performed with Hz-level residuals in the subsection of `Data processing', which experimentally justifies the negligence of the strain effect.

\section{Experimental methods}
\subsection{Diamond sample}
The NV centers reside in a bulk diamond whose top face is perpendicular to the [100] crystal axis and lateral faces are perpendicular to [110]. The nitrogen concentration of the diamond is less than 5 ppb and the abundance of $^{13}$C is at the natural level of 1.1$\%$. The diamond is irradiated using 10-MeV electrons to a total dose of $\sim$ 1.0 kgy. Four solid immersion lenses (SIL) are created on the diamond surface. Five NV centers are far away from these SILs, and two NV centers are inside one of SILs.

\subsection{Experimental Setup}
\begin{figure*}[t]\center
\includegraphics[width=1.\textwidth]{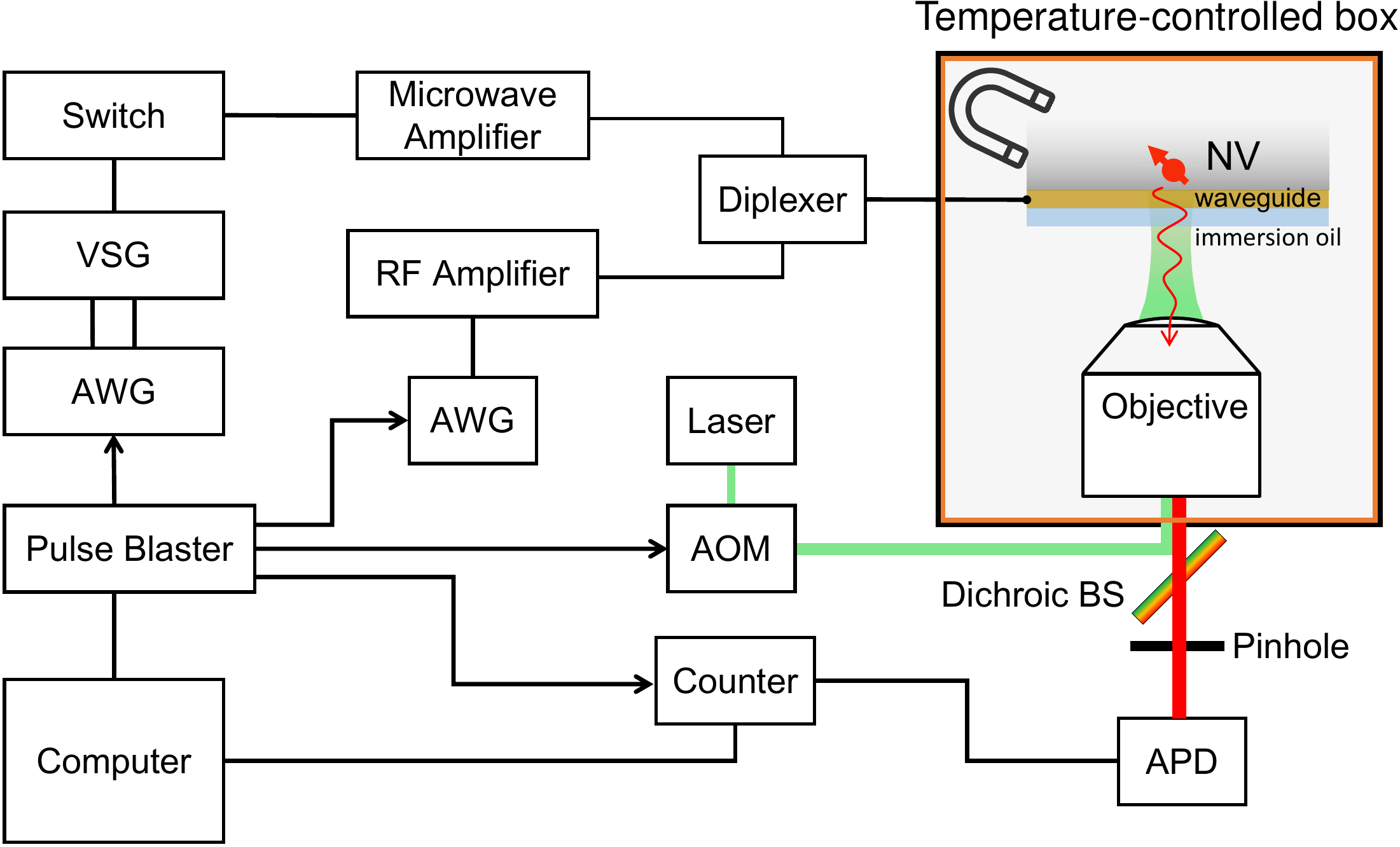}
\caption{\label{setup}\textbf{Simplified experimental setup.} The whole setup is automatically controlled by the pulse blaster and the computer for a convenient implementation. See detailed description in the subsection of Experimental Setup.
}
\end{figure*}

The simplified experimental setup is displayed in Fig. \ref{setup}. The diamond is mounted on a typical optically detected magnetic resonance confocal setup, synchronized with a microwave bridge by a multichannel pulse blaster (Spincore, PBESR-PRO-500). The 532-nm green laser for driving NV electron dynamics, and sideband fluorescence (650-800 nm) go through the same oil objective (Olympus, UPLSAPO 100XO, NA 1.40). To protect the NV center's negative state and longitudinal relaxation time against laser leakage effects, all laser beams pass twice through acousto-optic modulators (AOM) (Gooch $\&$ Housego, power leakage ratio $\sim$ 1/1,000) before they enter the objective. The fluorescence photons are collected by avalanche photodiodes (APD) (Perkin Elmer, SPCM-AQRH-14) with a counter card (National Instruments, 6612) embedded in the computer. The 4.3 GHz and 1.4 GHz microwave (MW) pulses for the manipulation of the NV three sublevels are generated from the microwave bridge, coupled with 0.1-10 MHz radio-frequency (RF) pulses for the nuclear spins via a diplexer, and fed together into the coplanar waveguide microstructure. The microwave bridge consists of an arbitrary waveform generator (AWG, Keysight, 81180A), a vector signal generator (VSG, Keysight, E8267D), a switch (PMI, P1T-DC40G-65-T-292FF-1NS-OPT18G-50OHM), and a high-power amplifier (Mini-Circuits, ZHL-25W-63+). The RF pulses are generated by another AWG (Keysight, 33522B) followed by another high-power amplifier (Mini-Circuits, LZY-22+). The external magnetic field ($\approx510$ G) is generated from a permanent magnet and aligned parallel to the NV axis through a three-dimensional positioning system. The positioning system, together with the platform holding the diamond and the objective, is placed inside a thermal insulation copper box. The temperature inside the copper box stabilizes down to a mK level through the PID feedback of the temperature controller (Stanford, PTC10). We elaborately design a good configuration for thermal insulation, which is easily neglected often under many circumstances and even more important than the PID feedback.

\subsection{Magnetic field stability}
\begin{figure*}[t]\center
\includegraphics[width=1.\textwidth]{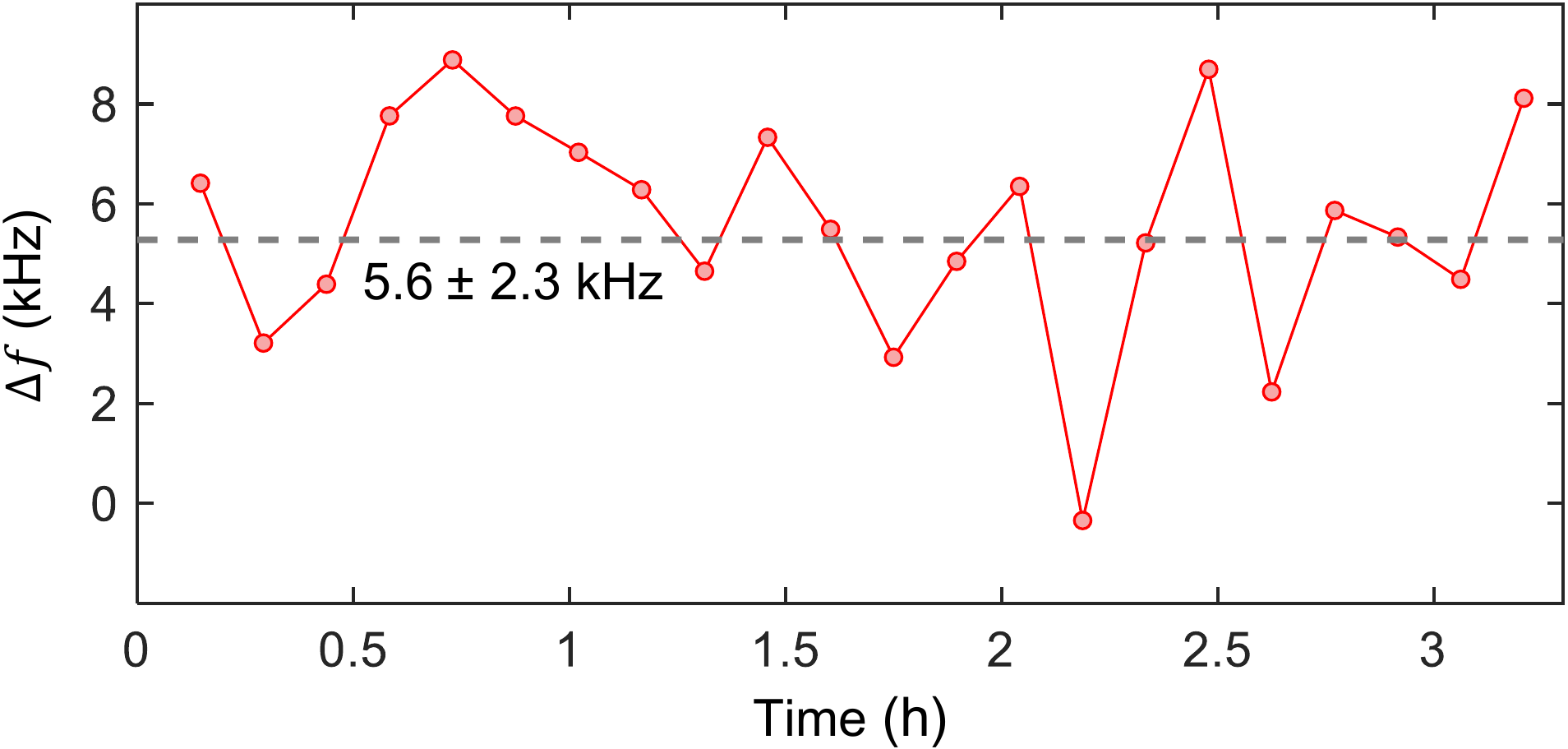}
\caption{\label{stability}\textbf{Magnetic field stability.} Time trace of the resonant dips of the sixth NV center inside the SIL over three hours. The mean value represented by the dashed line is 5.6 kHz with a frequency offset 4303.55 MHz and the fluctuation is 2.3 kHz (one standard deviation).
}
\end{figure*}

One of key indicators affecting the performance of our setup is magnetic field stability. It depends on many factors including temperature fluctuation, vibration and humidity fluctuation, and among them temperature fluctuation is the most important one. The temperature disturbs the magnetic field felt by the NV in two ways: first, it changes the magnetization of the magnets (0.12$\%$/\textcelsius\;for NdFeB magnets); second, it causes the stretch of materials and changes the position of the NV in the magnetic field. With an elaborately designed configuration for heat isolation and suitable PID parameters for feedback, the temperature fluctuation is suppressed down to $\sim$1 mK.

In order to estimate the fluctuation of the external magnetic field, we monitored the resonant dips of the sixth NV center inside the SIL over three hours as shown in Fig. \ref{stability}. The fluctuation of the resonant frequency is 2.3 kHz, which corresponds to a magnetic fluctuation below 0.1 \textmu T. As a matter of fact, the magnetic fluctuation from the external magnet should be better because the magnetic fluctuation contains the local magnetic fluctuation from $^{13}$C bath.

\section{Experimental results and data processing}
\subsection{Experimental results}
\begin{table*}\center
\includegraphics[width=1\textwidth]{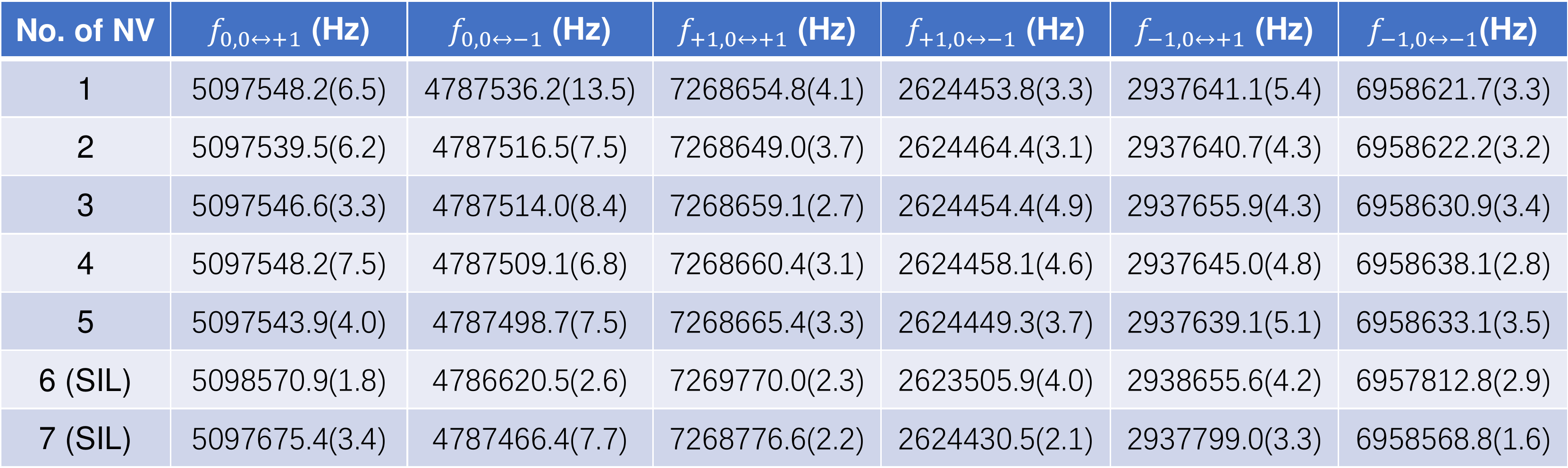}
\caption{\label{frequencies}\textbf{Transition frequencies of $^{14}$N nuclear spins for seven NV centers.} The first five NV centers are far away from SILs while the last two NV centers are inside a SIL. Find the definitions of the transition frequencies in Eq. (\ref{f_0to+1}) and (\ref{f_0to-1}).
}
\end{table*}

By adopting the scheme of Ramsey interferometry (see the description in the text), we obtained all transition frequencies of $^{14}$N nuclear spins for seven NV centers listed in Table \ref{frequencies}. Among them, the first five NV centers are far away from SILs which the last two NV centers are inside a SIL. The measured frequencies of two NV centers inside the SIL have better precisions than those of five NVs away from SILs because of higher luminescence rates for the NVs inside the SIL. The magnetic field for performing the measurements is $\approx509$ G except the sixth NV center $\approx512$ G. The values in parentheses stand for one standard deviation. 

\subsection{Data processing}
\begin{table*}\center
\includegraphics[width=1\textwidth]{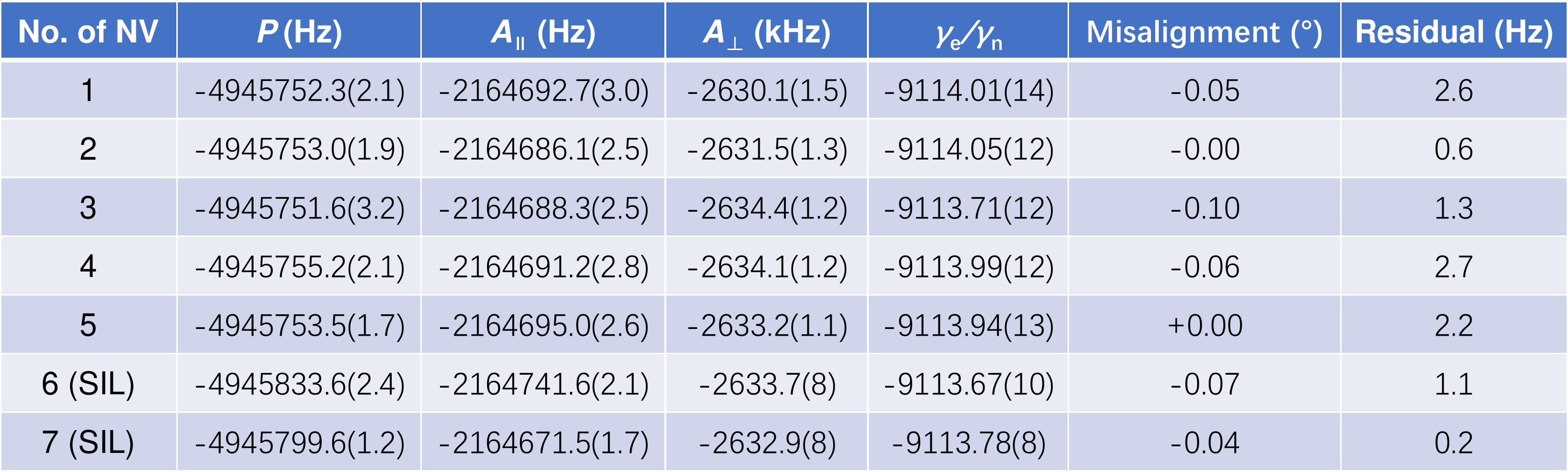}
\caption{\label{five}\textbf{Least squares regressions with five parameters.} The misalignment of the magnetic field with the NV axis is considered.
}
\end{table*}

\begin{table*}\center
\includegraphics[width=1\textwidth]{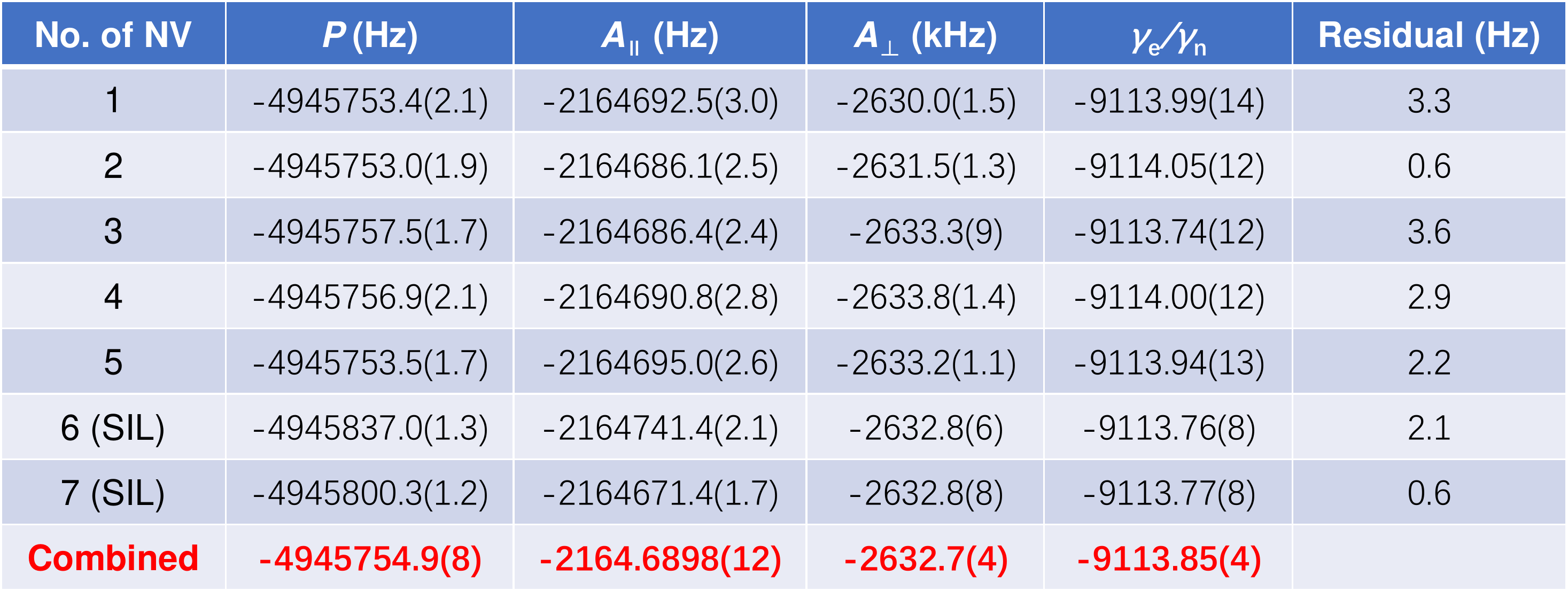}
\caption{\label{four}\textbf{Least squares regressions with four parameters.} The parameter, the misalignment of the magnetic field, is ignored here.
}
\end{table*}

As discussed in the text, we have five independent parameters and six measured transition frequencies for each NV center as listed in Table \ref{frequencies}. Since the number of the measured frequencies is greater than that of the parameters, we use the method of least squares regression to solve the parameters. We numerically acquired the optimal values for five parameters by finding the minimum of the weighted residual over the inverses of the errors for the measured frequencies. The weighted residual with the formulas $f_i(P,A_{\parallel}...)$ for each transition frequency and the corresponding measured value $f_i^m$ is given by
\begin{equation}
\rm{Res}=\sqrt{\sum_{i=1}^{6}\frac{1/err_i^2}{\sum_{i=1}^{6}1/{err_i^2}}\it(f_i(P,A_{\parallel}...)-f_i^m)^{\rm{2}}}
\label{residual}
\end{equation}
where $\rm{err}_{\it{i}}$ is the statistical error for the $i$th transition frequency. By adding a small variance to each frequency and monitoring the variances of the optimal values, we can numerically acquire the derivatives associating the five parameters and the six transition frequencies. Then the statistical errors can be obtained by the method of error propagation. The results are listed in Table \ref{five}. The residuals obtained by the least squares regression are smaller than the uncertainties of the measured transition frequencies so that the uncertainties of the misalignment can not be determined. Therefore, we abandoned the parameter of the misalignment and performed the process above again (Table \ref{four}). We found that the residuals of the four-parameter regression are still at the level of the uncertainties of the measured transition frequencies, which justifies the abandonment of the parameter of the misalignment. As shown in the row `Combined' of Table \ref{four}, the $^{14}$N quadrupole coupling $P$, the longitudinal component $A_{\parallel}$, the transverse component $A_{\perp}$ of the hyperfine interaction, and the ratio $\gamma_e/\gamma_n$ of the gyromagnetic ratio of the NV$^-$ electron to that of $^{14}$N, are  -4945754.9(8) Hz, -2164689.8(1.2) Hz, -2632.7(4) kHz and -9113.85(4), which are derived by weightedly averaging the available values over the statistical errors. The available values are those of the first five NV centers for $P$ and $A_{\parallel}$, and those of all seven NV centers for $A_{\perp}$ and $\gamma_e/\gamma_n$.

\subsection{Comparison with previous experiments}
\begin{table*}[b]\center
\includegraphics[width=1\textwidth]{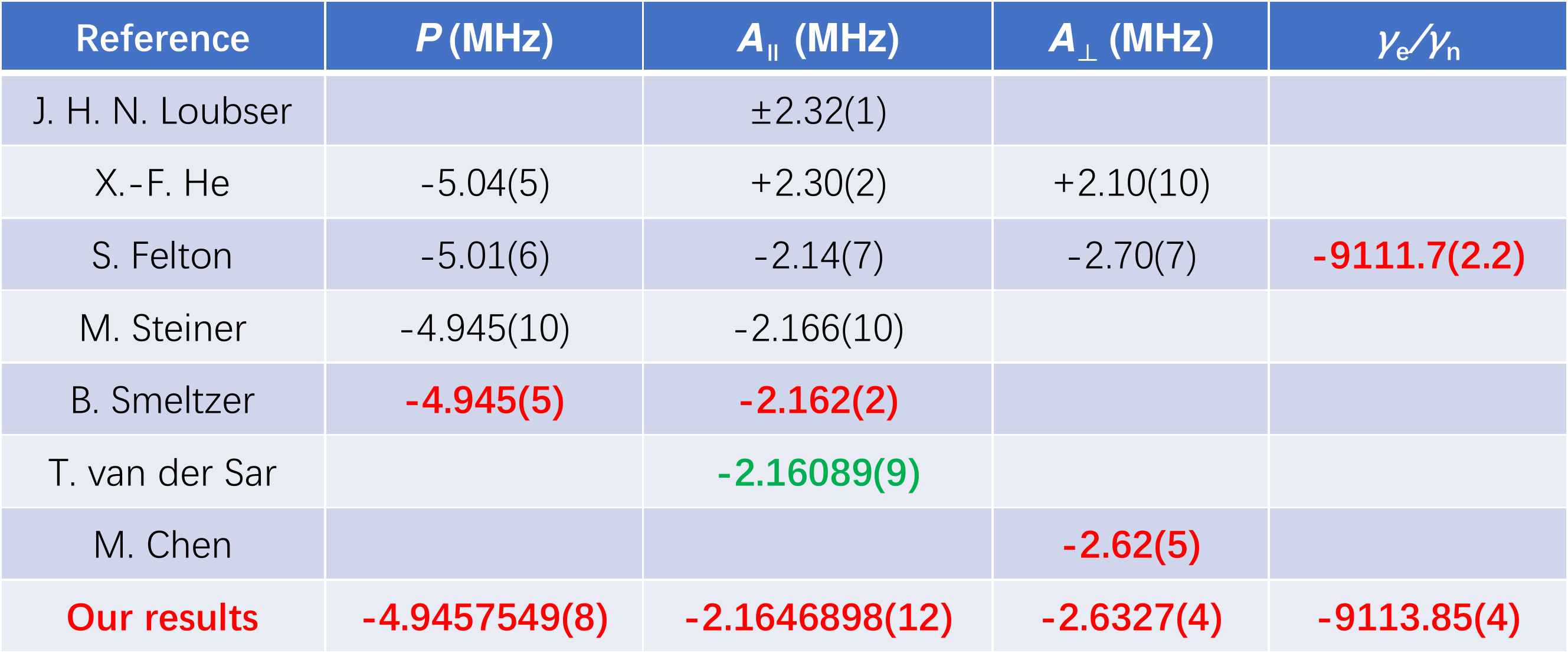}
\caption{\label{comparison}\textbf{Comparison with previous experiments.} The measurements of J. H. N. Loubser, X.-F. He, S. Felton \cite{gyro_2009}, M. Steiner \cite{enchancement_2010}, and B. Smeltzer \cite{nuclear_2009} are summarized in \cite{NV_2013}. The measurements of T. van der Sar and M. chen are presented in \cite{decoherence_protected_2012} and \cite{transverse_2015}. Both our results and the best results of previous measurements are marked in red and boldface for comparison.
}
\end{table*}

Our method can derive the values of all parameters involved in the NV$^-$-N Hamiltonian. Table \ref{comparison} collects the results of previous experiments as well as our results for comparison. The value marked in green has a fairly small statistical error, but it is accompanied by an enormous systematic error $\sim4$ kHz due to the overlook of $A_{\perp}$ in \cite{decoherence_protected_2012}. The value of -9111.7 is calculated by the gyromagnetic ratio of the NV$^-$ electron spin $\gamma_e\approx28.033(3)$ MHz/mT \cite{gyro_2009} and that of the $^{14}$N nuclear spin $\gamma_n\approx3.0766$ kHz/mT \cite{Philipp_2011}. The error of 2.2 is given by the deviation from our result. Compared with the best results of previous measurements, the precisions of our results are all higher by several orders of magnitude, nearly four orders of magnitude for $P$, more than three for $A_{\parallel}$, and nearly two for $A_{\perp}$ and $\gamma_e/\gamma_n$.

\section{Identity test on more NV centers}
\begin{figure*}[b]\center
\includegraphics[width=1. \textwidth]{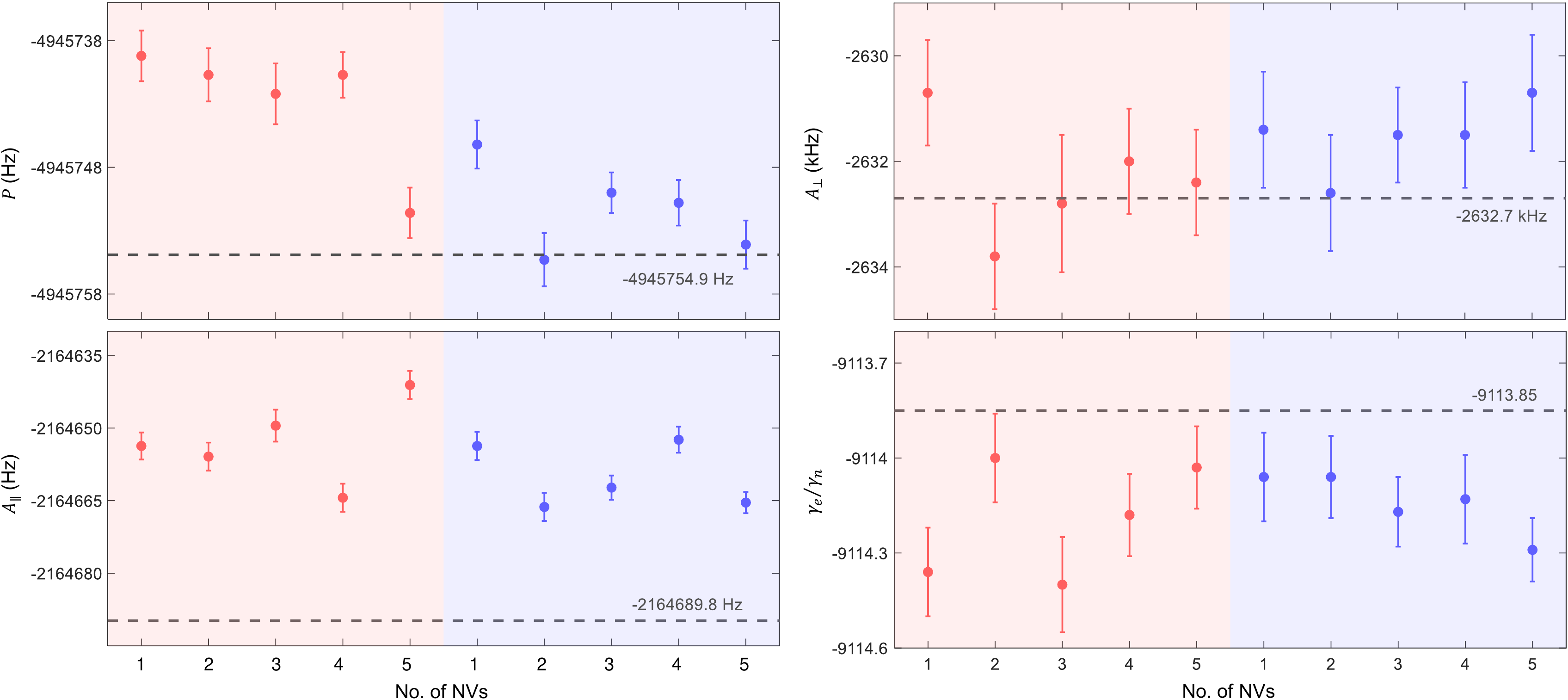}
\caption{\label{identity}\textbf{Identity test.} Like Fig. 3 in the main text, the identity test is further implemented on ten more NV centers away from SILs, among which the left five NVs stay in another district of the diamond sample in the main text and the right in another sample. The values represented by the dashed lines are the weighted averages of seven NV centers in the main text. All error bars represent one standard deviation.
}
\end{figure*}
Apart from the NV centers in the main text, we performed the identity test on ten more NV centers away from SILs, as shown in Fig. \ref{identity}.  Among them, the left five NVs stay in another area (20$\times$10 \textmu m$^2$) of the diamond sample in the main text, while the right five NVs are located within an 80$\times$20 \textmu m$^2$ area of another sample (prepared by the same procedures as the one in the main text). The parameters $P$ and $A_{\parallel}$ are still measured at Hz-precision level.

Unlike the five NV centers away from SILs, the NV centers here manifest unidentical in that most values are clearly above the dashed lines (the averages for the NV centers in the main text). It means that Hz-precision level is precise enough to differentiate general deep NV centers in different regions and diamond samples. Furthermore, unlike the NV centers inside a SIL in the main text, the deviations from the dashed lines are somewhat uniform, which signifies that the deviations are caused by the intrinsic strain in diamond. This kind of strain is somewhat uniform and may originate from sample growing or cutting. Especially, the first three NVs on the left are identical within errors, and are substantially different from the next two NVs. It is likely that the region of these five NV centers is near the border of different strain districts. Besides, the values of $\gamma_e/\gamma_n$ are all below the dashed line, which means that it is also sensitive enough to be used as an indicator to distinguish NV centers.

In the future, the phenomenon above could be related to the strain shifts of NV excited states at cryogenic temperatures, and our method (room temperature) may provide some information for the strain shifts at cryogenic temperatures \cite{cryo_strain_2009, cryo_strain_2011, cryo_strain_2016}. Furthermore, we can carry out a more comprehensive investigation on the strain distribution in diamond and even image it under ambient conditions. Last but not least, a distribution of tens of Hz nearly does not affect the coherence time of the nuclear spin $T_2^* \sim 10$ ms and thus will not lower the performance of the proposed atomic-like clock in Eq. (8) in the main text.

\section{The effect of the impurities surrounding the NV center}
The impurities surrounding the NV center may shift $P$ and $A_{\parallel}$ through the electric field effect, of which the most abundant one is nitrogen. The nitrogen concentration of the diamond we used here is less than 5 ppb (i.e., the average distance between two nitrogen atoms is more than 100 nm), and thus has no effect on $P$ and $A_{\parallel}$. As a matter of fact, the ensemble NV centers in the diamond with a nitrogen density of $\sim1$ ppm could have a dephasing time up to 10 \textmu s under single-quantum Ramsey sequence \cite{dephasing_2018}. It means that the zero-field splitting $D$ has a Gaussian distribution with a standard deviation of $\sim20$ kHz, which is induced by crystal strain gradients. The two NV centers inside a SIL have a zero-field splitting deviation of $\sim300$ kHz, compared with the NV centers far away from SILs. From our results, these two NV centers differ by tens of Hz in terms of $P$ and $A_{\parallel}$. Therefore, it can be inferred by combining these facts that $P$ and $A_{\parallel}$ may have a distribution of several Hz in an ensemble NV sample with a nitrogen density of $\sim1$ ppm. The deviation at this level has no effect on the coherence time of the nuclear spin $T_2^*\sim10$ ms and thus will not lower the performance of the proposed atomic-like clock in Eq. (8) in the main text.

\end{CJK*}